\documentstyle[11pt,epsf]{article}
\def\beq{\begin{eqnarray}}
\def\eeq{\end{eqnarray}}
\def\bea{\begin{eqnarray*}}
\def\eea{\end{eqnarray*}}

\def\centeron#1#2{{\setbox0=\hbox{#1}\setbox1=\hbox{#2}\ifdim
\wd1>\wd0\kern.5\wd1\kern-.5\wd0\fi
\copy0\kern-.5\wd0\kern-.5\wd1\copy1\ifdim\wd0>\wd1
\kern.5\wd0\kern-.5\wd1\fi}}
\def\ltap{\;\centeron{\raise.35ex\hbox{$<$}}{\lower.65ex\hbox{$\sim$}}\;}
\def\gtap{\;\centeron{\raise.35ex\hbox{$>$}}{\lower.65ex\hbox{$\sim$}}\;}

\def\singleandthirdspaced{\baselineskip=\normalbaselineskip\multiply
    \baselineskip by 130\divide\baselineskip by 100}

\newcommand{\newc}{\newcommand}
\newc{\qbar}{{\overline q}}
\newc{\Kahler}{K\"ahler }
\newc{\deltaGS}{\delta_{\rm GS}}
\begin{document}
\begin{titlepage}
\begin{flushright}
{\large hep-th/0412129 \\ SCIPP-2004/54\\
}
\end{flushright}

\vskip 1.2cm

\begin{center}

{\LARGE\bf Landskepticism {\it or} \\ Why Effective Potentials Don't
Count String Models}

\vskip 1.4cm

{\large  T. Banks}
\\
\vskip 0.4cm
%{\it $^a$Stanford Linear Accelerator Center,
%     Stanford CA 94309} \\
{\it Santa Cruz Institute for Particle Physics,
     Santa Cruz CA 95064  } \\
     {\it NHETC, Rutgers U., Piscataway, NJ 08854}
%{\it $^c$Physics Department,
%     University of California,
%     Santa Cruz CA 95064  }

\vskip 4pt

\vskip 1.5cm

\begin{abstract}
This paper is a synthesis of talks I gave at the Cargese Workshop in
June 2004 and the Munich Conference on Superstring Vacua in November
2004.  I present arguments which show that the landscape of string
theory is not a well established feature of the theory, as well as a
brief discussion of the phenomenological prospects of the landscape
and the use of the anthropic principle.

\end{abstract}

\end{center}

\vskip 1.0 cm

\end{titlepage}
\setcounter{footnote}{0} \setcounter{page}{2}
\setcounter{section}{0} \setcounter{subsection}{0}
\setcounter{subsubsection}{0}

%%%%%%%%%%%%%%%%%%%%%%%%%%%%%%%%%%%%%%%%%%%
%%%%%%%%%%%%%%%%%%%%%%%%%%%%
\singleandthirdspaced

\section{\bf Introduction}

This paper is a synthesis of my summary remarks at the Cargese
School, and my talk at the Munich conference. Its main message is
that the landscape of string theory is far from an established fact.
I point out that a hypothetical rigorous formulation of the
landscape will require us to construct mathematical models of both
the Big Bang and Eternal inflation.   I emphasize that without such
a rigorous formulation we are unlikely to be able to calculate the
properties of the landscape with sufficient accuracy to make
comparisons with experiment.   I briefly review the phenomenological
difficulties of the landscape idea, and make some comments about the
anthropic principle.

The reader should be warned in advance that my personal bottom line
on this subject is that the Landscape probably does not exist.
However, I have tried to give it the maximal benefit of the doubt.
Thus in many sections of this paper I will be trying as hard as
possible to think of what the landscape might mean in a real theory
of quantum gravity, which has the sort of solidity we associate with
perturbative string theory, M-theory on Calabi-Yau manifolds times a
torus, AdS/CFT, or Type 0 string theory in $1 + 1$ dimensional
linear dilaton space-times.  I hope that the encouraging tone of
some of these sections will not distract the reader from my main
message.

One other caveat.  I use the term FRW cosmology as a short-hand for
a Friedman-Robertson-Walker cosmology which does not undergo eternal
acceleration.

\section{\bf Effective actions in string theory}

Part of the confusion in the debate about the landscape is the
conflation of two notions of effective action, both of which occur
in quantum field theory.   The first is the 1PI effective action,
which is an exact summary of the entire content of a field theory.
Knowledge of it enables us to construct all of the correlation
functions of the theory, in any of its vacuum states.   In
perturbation theory, we can compute the 1PI action around any vacuum
state, and get the same result.   It is important to realize that
there is NO known analog of the 1PI action in string theory, which
applies to all solutions of the theory, and enables us to pass from
one to the other by ``shifting the field".

The other concept of effective action in field theory is the low
energy or Wilsonian action.  This is defined, either in a single
vacuum, or in a set of quasi-degenerate vacua whose energy density
differences (as well as the heights of the barriers between them)
are small compared to some cutoff scale.  The Wilsonian action only
contains degrees of freedom whose fluctuations are significant at
these low energy scales.  It is important to avoid using it when the
conditions of its validity are violated.  For example, in minimally
SUSic QCD with $N_F < N_C - 1$, the low energy degrees of freedom
consist of a meson superfield $M$ and one can compute the exact low
energy superpotential of $M$.  The low energy Kahler potential is
canonical.  If one uses this low energy Lagrangian to compute the
energy density of states with a given expectation value of $M$, this
Lagrangian gives a divergent answer at the symmetric point of
vanishing eigenvalue.   This is not the correct physical answer. The
true Kahler potential (in the 1PI sense) of $M$ is modified at the
origin of moduli space, and the energy density there is really of
order the QCD scale.   We should stop paying attention to the
predictions of the Wilsonian action when they are outside its range
of validity.

In string theory, the effective actions we compute are analogous to
Wilsonian actions, but their range of validity is even more
constrained.   In particular, the stringy derivation of the
effective action views it as a tool for calculating boundary
correlation functions in a {\it fixed asymptotic space-time
background}. We tend to forget this because, particularly in
situations with a lot of SUSY, the leading low energy term in the
effective action is independent of the background.   This fosters
the illusion that different backgrounds can be viewed as {\it vacuum
states of the same theory}.  In fact, as emphasized in
\cite{hetero}, the italicized phrase is borrowed from quantum field
theory, and refers to concepts that depend entirely on the
separation between IR and UV physics of that formalism.  In string
theory/quantum gravity, UV and IR physics are much more intimately
entangled, and the concept of different vacuum states of the same
underlying string theory Hamiltonian is much more circumscribed.

When we have a continuous moduli space of super-Poincare invariant
S-matrices, we can do experiments\cite{isovac} at one value of the
moduli, which are sensitive to the S-matrix at other values of the
moduli\footnote{These experiments are much more difficult than they
would be in a SUSic quantum field theory, without gravity.}. Note
however, that the only Hamiltonian form we have for such models is
in light cone frame, where different values of the moduli correspond
to different Hamiltonians. Similarly, a moduli space of correlation
functions on the boundary of Anti-deSitter space, corresponds to a
one parameter set of different Hamiltonians, rather than different
superselection sectors of the same Hamiltonian.   The notion of
different vacua of the same theory, in any of the senses that this
is meant in quantum field theory, is simply not applicable to
theories of quantum gravity, beyond the very limited context of
continuous moduli spaces of Super-Poincare invariant S-matrices.

The recognition that changes in background asymptotics correspond to
changes in the Hamiltonian, rather than changes of superselection
sectors for a given Hamiltonian goes back to \cite{ss}, and has
become commonplace with the advent of AdS/CFT. Changes in
non-normalizable modes of bulk fields\footnote{Note that if we want
to make finite rather than infinitesimal changes we must restrict
our attention to Breitenlohner-Freedman allowed tachyon fields.} add
relevant terms to the Hamiltonian. Changes in the (negative)
cosmological constant correspond to changes in the fixed point which
defines the boundary CFT, and thus to a completely different set of
high energy degrees of freedom.

These facts lead us\footnote{well, at least they lead me} to be
suspicious of attempts to find new string theory models by
patching together an effective potential using the degrees of
freedom of {\it e.g.} Type II string theory in flat space-time.
Indeed, once we include gravitational effects, the low energy
action itself gives us reason to be suspicious of meta-stable de
Sitter vacua constructed in this way.

In my opinion, {\it the concept of an effective potential on moduli
space as a tool for finding string models of gravity, is a snare and
a delusion, fostered by wishful thinking, and without regard to the
actual evidence in front of us. There is no evidence for this
concept in solid string theory calculations, and lots of evidence
against it\footnote{Those who think these judgements too harsh,
should recall that I've spent a considerable fraction of my career
writing about effective potentials on moduli space.  I've also spent
the last four years trying to draw the community's attention to what
I believe is the true situation - with frustratingly little effect.
Perhaps some over the top rhetoric is in order.}. As I will explain
below, the utility of the effective potential concept in low energy
effective non-gravitational physics is not an argument for its
validity in string theory.  Gravitational low energy effective field
theory shows us the limitations of the effective potential concept,
and is a good tool for demonstrating that different solutions of a
low energy effective action do not define different states of the
same quantum theory unless they have the same space-time
asymptotics. Rigorous quantum string theory constructions, including
$c \leq 1$ matrix models, Matrix Theory, and AdS/CFT, confirm these
conclusions.}

The fact that the effective actions we compute in string theory are
not analogs of a 1PI action is easily confirmed by calculations we
can do in well defined string theories.  We can consider, among many
examples, the Type IIB supergravity solutions $M^{1,d} \times K^{9 -
d}$ and $AdS_5 \times S^5$, where $K$ is a SUSY preserving compact
manifold.  The effective actions appropriate to these different
examples, depend on different kinds of variables. Different $K$ give
rise to different moduli fields.  The constant value of the dilaton
is a variable in the Minkowski actions, while it is fixed and
corresponds to a parameter in the quantum Hamiltonian in AdS.  The
AdS space has singleton fields, which have no Minkowski analog.
Higher order corrections to these actions involve virtual
contributions from states that only exist for some choices of $K$.
Even the terms involving fields common to all the different
non-compact space-times will, in general, have different
coefficients in the different backgrounds (except for terms which
are completely determined by symmetries).   There is no way to
change $K$ from {\it e.g.} a torus to a Calabi-Yau manifold, or to
change Minkowski into AdS, by ``shifting a field in the quantum
effective action".   And, indeed, these actions are computed, by
calculating boundary correlation functions with a fixed asymptotic
boundary geometry, and finding an action functional whose tree level
solutions match the low energy expansion of these boundary
correlators.

\subsection{The effective potential with and without gravity}

 Consider a simple scalar field theory model with a potential that
 has one positive energy minimum and one at exactly zero energy.  We
 can imagine the zero energy minimum to be related to an exactly
 supersymmetric string theory vacuum state.   Let the scale of all
 parts of the potential in the region of the two minima be small
 compared to both the string and Planck scales.   It is then a
 correct deduction from low energy effective field theory, that a
 subclass of solutions of this Lagrangian, including some in which
 the field is homogeneous and equal to its false vacuum value
 over a large region of space, and asymptotes to the true vacuum at
 infinity, can be studied without worrying about gravity.   However,
 {\it it is not true that all solutions whose energy density is
 small compared to the Planck scale can be studied in this way.}
 We can certainly excite a bubble of the meta-stable minimum, and watch it
 decay, but if the size of the bubble gets too large, we run into problems.

In order to exhibit a meta-stable dS space, we have to excite a
region of order the putative Hubble radius into the metastable
minimum. Old results of Guth and Farhi\cite{gf} show that the
external observer can never verify the existence of the inflating
region.   A black hole forms around it.  Any observer in the
asymptotically flat region, who tries to jump into the black hole to
find inflation, first encounters a singularity. So effective field
theory tells us that, given a well defined, Poincare invariant
S-matrix for gravity, we can find and explore meta-stable positive
energy density minima of an effective potential, but not the
meta-stable dS spaces that these minima have been thought to imply.
That is, the Guth-Farhi results suggest that {\it the stable
asymptotically flat vacuum state does not have excitations which
correspond to metastable dS vacua, even when it has an effective
potential with positive energy meta-stable minima}. Rather, it has
excitations in which fields are excited into metastable minima only
over regions small compared to the Hubble radius at those minima.
The attempt to create larger regions succeeds only in creating black
holes.  Similar remarks apply to stable AdS vacua, whether they are
supersymmetric or not.

We see that in gravitational theories, the criterion for the
validity of non-gravitational effective field theory reasoning
depends on more than just the value of the energy density in Planck
units.  When the Schwarzschild radius of a region exceeds its
physical size (in the approximation in which gravity is neglected),
a black hole forms.  Effective field theory remains valid outside
the black hole horizon (if it is large enough), but not inside.  In
the above example, no external observer can probe the putative dS
region, without first encountering a singularity.

We also see another example of the principle that the solutions of
the same effective equations of motion may not reside in the same
quantum theory.  This is in stark contrast to non-gravitational
quantum field theory with the same potential.  There, the
meta-stable minimum of the potential represents an unstable Lorentz
invariant state of the Hamiltonian whose ground state is the zero
vacuum energy state.  We can creat arbitrarily large regions where
the field lies in its meta-stable minimum.

The results of Coleman and De Lucia\cite{cdl}, on vacuum tunneling
in the presence of gravity, give us a sort of converse to these
observations. Given the same effective Lagrangian we used in the
previous paragraph, we can assume the existence of the meta-stable
dS space, and ask what it decays into.   Here there is a surprise,
particularly for those who constantly repeat the mantra ``dS space
decays into flat space".   In fact, the analytic continuation of the
CDL instanton is a negatively curved Friedmann Robertson Walker
cosmology, which (if the potential has a stable zero cosmological
constant minimum), is asymptotically matter dominated.   Although it
locally resembles flat space on slices of large cosmological time,
its global structure is completely different.  In particular, an
attempt to set up an asymptotically Minkowskian coordinate system,
starting from the local Minkowski frame of some late time observer,
inevitably penetrates into regions where the energy density is high
(the energy density is constant on slices of constant negative
spatial curvature, and is of order the energy density of the false
vacuum at early FRW times) .   Another indication that the local
point of view is misleading is that if we make the vacuum energy
even slightly negative, the asymptotic state is radically altered:
we have a Big Crunch rather than a small perturbation of Minkowski
space.  Models of quantum gravity with AdS asymptotics cannot have
meta-stable de Sitter (or Minkowski) excitations.

Thus, both analysis of creation and decay of meta-stable dS states,
suggests that if a potential has a stable minimum with vanishing
cosmological constant, and another with positive energy density, the
Minkowski solution and the meta-stable dS solution are simply not
part of the same theory. There remains a possibility of the
existence of a theory with a stable, matter dominated, FRW
cosmological solution with a meta-stable dS excitation.  The problem
with this is that it is very unlikely that we can make a reliable
exploration of this scenario within the realm of low energy
effective field theory.

The CDL instanton solution is non-singular. The $a=0$ point of the
FRW coordinate system is just a coordinate singularity marking the
boundary between the FRW region and a region of the space-time which
continues to inflate. However, arbitrary homogeneous perturbations
of the CDL solution have curvature singularities. Further, there is
a large class of localized perturbations which evolve to Big Crunch
singularities, rather than passing smoothly through the $a=0$ point.
For example, consider a localized perturbation on some hyperbolic
time slice a finite proper time prior to $a=0$.  Let it be
homogeneous in a large enough region, so that signals from its
inhomogeneous tail cannot propagate to the $a=0$ point.   Then we
will have a singularity.  Below we will see an argument that
singularities are associated with generic initial\footnote{In
mentioning initial data, we are referring to the time symmetric
Lorentzian CDL instanton.} and final data in this geometry.

If we re-examine the Guth-Farhi argument in the FRW context, we see
that it continues to hold until we go back in time to a point where
the cosmic energy density is of order the barrier to the meta-stable
minimum.   At high enough energy density, there are FRW solutions in
which the field classically evolves into the meta-stable minimum. It
will then decay by tunneling, into the FRW continuation of the CDL
instanton. Generic FRW solutions (including arbitrary homogeneous
perturbations of the CDL instanton) have curvature singularities at
a finite cosmic time in the past. To establish their existence as
genuine theories of quantum gravity one must go beyond effective
field theory, and probably beyond perturbative string theory.

Freivogel and Susskind\cite{susspriv} have suggested a scattering
theory in which asymptotic states are associated with incoming and
outgoing wave perturbations of the nonsingular, time symmetric
Lorentzian continuations of the, CDL instantons for the various
meta-stable vacua of string theory. They claim that in this
framework, the breakdown of effective field theory is avoided, as
long as the effective potential is everywhere smaller than Planck
scale.   I find this suggestion interesting, but it is not based on
reliable calculations, and there is a strong indication that it does
not work. If one considers black holes with radius larger than the
dS radius, formed in the remote past of the FRW part of the time
symmetric Lorentzian CDL geometry, it is hard to see how these
solutions asymptote to the future CDL geometry without encountering
a singularity. Near the $a = 0$ point where the FRW coordinates on
the CDL solution become singular, most of the space-time is
isometric to dS space at its minimal radius. If we have formed a
black hole in the past, with radius much larger than this, then the
entire spacetime must end in a singularity.

Thus, I would claim that unlike asymptotically flat or AdS
space-times, we have no reliable effective field theory argument
that there are an infinite number of states in space-times that
asymptote to the time symmetric Lorentzian CDL instanton.  An
infinite number of states is a minimal requirement for the existence
of a quantum theory that can make precise mathematical predictions,
which can be self consistently measured in the theory.  We must
understand the nature of the singularities in these perturbations of
the CDL instanton geometry before we can conclude that the framework
makes sense.

I think it is more likely, that meta-stable dS vacua
exist\footnote{if they exist at all!} only in the context of a Big
Bang cosmology. If we consider the problem of accessing a
meta-stable dS minimum of an effective potential, at times when the
cosmic energy density is of order the barrier height of the
potential, then the Guth-Farhi problem does not appear to exist.
Starting from a Big Bang singularity, one can find homogeneous
solutions where a scalar field wanders over its potential surface at
a time when the energy is higher than the barriers between minima,
and then settles in to a meta-stable minimum with positive
cosmological constant.  One must understand the Big Bang to make a
reliable theory of such a situation, but apart from that the
solution is non-singular.   In particular, the problem of large
black holes in the initial state, is not present for this situation.
The difficulties are all associated with understanding the Big Bang
singularity, and with Eternal Inflation, which I will discuss below.
However, I will also point out below that generic final states in
the CDL instanton geometry cannot have evolved from a meta-stable dS
tunneling event - only a finite number of states are compatible with
this history.

To summarize, it is clear from semi-classical calculations alone,
that the concept of a vacuum state associated with a point in scalar
field space is not a valid one in theories of quantum gravity.  A
given low energy effective field theory may have different solutions
which do not have anything to do with each other in the quantum
theory.   One solution may be a classical approximation to a well
defined quantum theory, while the other is not.  It seems likely
that the context in which we will have to investigate the existence
or non-existence of the landscape is Big Bang cosmology.   This is
the only situation in which we can ``reliably" construct a universe
which gets stuck in a meta-stable dS minimum.

Thus, I claim that if string theory really has a multitude of
meta-stable dS states, then the exact theory into which they fit is
a theory of a Big Bang universe which temporarily gets stuck, with
some probability, in each of these states.  This is ultimately
followed by decay to negatively curved FRW universes.  These FRW
universes have four infinite dimensions and $6$ or $7$ large compact
dimensions, which are expanding to infinity.   It is clear that the
probability for finding a particular dS vacuum is partly determined
by the density matrix at the Big Bang and not just by counting
arguments. In the next subsection I will describe existing proposals
for the cosmological distribution of vacua.   My main point here is
that the nature of the Big Bang will have to be addressed before we
can hope to understand the correct statistics of what are called
{\it stringy vacua}.

\subsection{Eternal inflation}

The string landscape seems to fit in well with older ideas which go
under the name of {\it eternal inflation}, {\it the self reproducing
inflationary universe}, {\it etc.}.   The simplest model which
exhibits this sort of behavior is one with a single scalar field
with two minima, one with positive vacuum energy and the other with
vanishing energy.   One considers the expanding branch of the
meta-stable dS universe with the field in the false minimum.   In
the quantum field theory approximation this seems to produce an
ever-expanding region of space and one allows the dS space to decay
by CDL bubble formation, independently in each horizon volume. In
the eternal inflation picture, one tries to interpret the result as
a single classical space-time. One obtains a Penrose diagram with a
future space-like boundary.  The future space-like boundary is
fractal, with regions corresponding to singularities\footnote{The
singular regions correspond to decays to parts of the potential
where the vacuum energy is negative. They would exist in the
landscape context, but not in the simple model we are discussing.}
as well as FRW asymptotics, interspersed in a causally disconnected
way.  In the landscape there will be many different singular regions
of the boundary, as well as many different FRW regions.  Advocates
of the landscape/eternal inflation picture then make the analogy to
maps of the observable universe, with different causally
disconnected regions being the analogs of different planets. Physics
it is said, depends on ``where you live" and organisms like
ourselves can only live in certain regions of the map.   A key
difference between different minima of the effective potential in
eternal inflation, and different planets is that {\it we cannot,
even in principle, communicate with causally disconnected regions of
the universe.}

How is one to interpret such a picture in terms of conventional
quantum mechanics?   I believe that the fundamental clue comes from
the principle of Black Hole Complementarity\cite{tHsuss}. Black
holes also present us with two regions of space-time which are
causally disconnected.  Hawking showed long ago that this was an
artifact of the semi-classical approximation, and that black holes
return their energy to the external space-time in which they are
embedded.   If we assume that the region behind the horizon has
independent degrees of freedom, commuting with those in the external
space, then we are confronted by the information loss paradox.

String theorists have believed for some years now, that this is
not the case.   The principle of Black Hole Complementarity is the
statement that the observables behind the horizon do not commute
with those in the external space-time.   For a large black hole,
(and for a long but finite time as measured by the infalling
observer), these two sets of observables are both individually
well described by semiclassical approximations, but the two
descriptions are not compatible with each other.

Fischler and I tried to relate this principle to the {\it Problem of
Time}\cite{mcosmo}.   In the semi-classical quantization of gravity
one attempts to solve the Wheeler-DeWitt equation $${{\cal H}\Psi =
0} $$ with an ansatz $${\Psi = e^{i S} \chi (t, \phi),} $$ where $S$
is the action of some classical space-time background solution, and
$\chi$ is the wave functional of a quantum field theory in this
space-time $${i\partial_t \chi = H(t) \chi}.$$ The (generally time
dependent) Hamiltonian $H(t)$ depends both on the choice of
classical background, and on the particular time slicing chosen for
that background.   For example, for the Schwarzschild background we
could choose $H_{Sch}$, the Schwarzschild Hamiltonian, or some time
dependent $H(t)$ where $t$ is the proper time of a family of
in-falling observers.   Even ignoring subtle questions of whether
these two Hamiltonian evolutions act on the same Hilbert space, it
is clear that they are different and that $[H(t) , H_{Schw}] \neq 0$
at {\it any} time $t$. It is therefore not surprising that the
semi-classical observables of different observers do not commute
with each other.

Given this description of black holes, it is natural to conjecture
that a similar phenomenon occurs for any space-time with horizons.
In \cite{mcosmo} this was called Cosmological Complementarity for
asymptotically dS spaces.  E. Verlinde has suggested the name
Observer Complementarity for the general case.

If we apply this logic to Eternal Inflation, we obtain a picture
quite different from the original description of these space-times.
We simply associate a single Hilbert space and many different
(generally time dependent) Hamiltonians to the fractal Penrose
diagram. Each Hamiltonian is associated with the causal patch of a
given observer. Mathematically, the situation can be equally well
described by saying we have a collection of different theories of
the universe. There is a philosophical cachet, associated with the
phrase, ``physics depends on where you live in the multiverse",
which is absent from this alternative way of describing the physics.

In the formalism described in \cite{susspriv} this is almost
precisely what is conjectured.   For each meta-stable dS point $L$
in the landscape, which can decay into the Dine-Seiberg region of
moduli space, and for each (typically 10 or 11 dimensional)
Super-Poincare invariant solution $V$ of string theory ``into which
$L$ can decay" there is a different unitary S-matrix
$S_{L,V}$\footnote{We can also consider initial and final states
corresponding to different CDL instantons.   In\cite{susspriv} these
are claimed to be different gauge copies of the same information in
the S matrices.  I will mention a different interpretation below.}.
It is claimed that each of these S-matrices contains all the physics
of the landscape, because there is a canonical way to compute the
unitary equivalence $U$ in the formula $S_{L,V} = U_{L,V,L^{\prime},
V^{\prime}} S_{L,V} U_{L,V,L^{\prime}, V^{\prime}}^{\dagger}$. The
statement that there is a theory of eternal inflation in which all
of the points $L$ are meta-stable states is really the statement
that the theory contains a canonical algorithm for computing the
``gauge transformations" $U$\footnote{For the moment, no approximate
statement of what this algorithm is has been proposed. I would
conjecture that, if the formalism makes sense, the transition
amplitudes between two different FRW spacetimes, mentioned in the
previous footnote, provide the algorithm for calculating the $U$
mappings. }. The authors of \cite{susspriv} claim that all of the
meta-stable dS states will show up as resonances in every S-matrix,
$S_{V,L}$. This claim is plausible if the S-matrices are indeed
related by unitary conjugation.  The spectrum of the S-matrix is
then gauge invariant. In ordinary scattering theory, time delays,
which are related to resonance lifetimes, are related to the
spectrum of the S-matrix. In the eternal inflation context, there is
no universal notion of time for the different asymptotic states, so
more work is necessary to understand these concepts.

If indeed the information about each meta-stable dS vacuum can be
extracted from the spectral density of a given S-matrix $S_{V,L}$,
and if we can find a reliable framework, for defining and
calculating these S-matrices, then the landscape will have a
mathematical definition.   From a practical point of view however,
we just say that string theory gives us an algorithm for
constructing models of the world (in the landscape context this
means choosing particular stable or meta-stable minima) which is not
unique and that we are trying to use data to constrain which model
we choose. In the penultimate section, where I summarize the
phenomenological difficulties of the landscape, I will describe the
situation in the language of the approximate Hamiltonian of a given
metastable dS observer, rather than that of eternal inflation.

I have emphasized the problems with this S-matrix point of view and
suggested the alternative notion that the landscape could only make
sense in the context of Big Bang Cosmology. One must thus understand
how to describe the initial states.  There are two possibilities,
either there is some principle which picks out a fixed initial
state\footnote{{\it e.g.} the Hartle Hawking Wave Function of the
Universe.} at the Big Bang, or there is a generalized S-matrix in
which we relate a particular state at the Big Bang to a particular
linear combination of final scattering states in one of the FRW
backgrounds defined by a decaying meta-stable dS space. In a manner
analogous to Freivogel and Susskind, one would conjecture that the
descriptions in terms of different future FRW backgrounds are
unitarily equivalent to each other, by unitary transformations which
do not respect locality.   Since we are unlikely to have much
control over the initial conditions at the Big Bang, one should
choose the in-state at the Big Bang to be a high entropy density
matrix. Thus, the practical difference between the Hartle-Hawking
(S-vector) and S-matrix proposals is the entropy of the initial
state.

I know of two proposals for the initial density matrix at the Big
Bang, which might lead to a set of cosmological selection rules for
meta-stable points in the landscape.  The first, {\it modular
cosmology}\cite{moorehorne} postulates an early era in which the
universe can be described semi-classically, but the potential on
moduli space is smaller than the total energy density.   The metric
on moduli space then provides a finite volume measure. Furthermore,
the motion of the moduli is chaotic.   These facts suggest that the
probability of finding the universe in a given meta-stable minimum
of the potential is the volume of the basin of attraction of that
minimum, divided by the volume of moduli space.

Holographic cosmology\cite{tbf} gives an alternative view of the
initial state of the universe, as a ``dense black hole fluid" where
standard notions of local field theory do not apply.   The model
contains two phenomenological parameters, which govern the
transition between this phase of the universe and a normal phase in
which the field theory description is valid.   There are indications
that the transition occurs at an energy density well below the
unification scale of standard model couplings.   We might then
expect a transition directly into a state with most of the moduli
frozen.   In order for the model to provide an adequate account of
the fluctuations in the CMB, one must have at least one ``active"
modulus at these low energies, which can provide for a modest number
of e-folds of inflation.  In such a model, minima of the potential
on moduli space, with energy higher than the scale at which a field
theoretic description of the universe is possible, cannot make any
sense.   At best a small class of low energy minima could be
compatible with holographic cosmology\footnote{The potentials
calculated in the landscape have no indication of a cut-off at
energy scales far below the unification scale.   This suggests that
the two theoretical frameworks are not compatible, but I am trying
to avoid jumping to conclusions.}.  If holographic cosmology is
compatible with landscape ideas, the probability of accessing a
particular minimum will be determined by quantum gravitational
considerations, far removed from effective field theory.   In this
framework, the dense black hole fluid is stable and is the most
probable state of the universe.   A normal universe like our own is
determined by a somewhat improbable initial condition, but one
expects the maximum entropy initial state that does not collapse to
a dense black hole fluid.   The survival probability of a given
normal state depends on both the properties of the black hole fluid
and the low energy physics of the normal state, so the determination
of the most probable meta-stable minimum would be a complicated
quantum gravitational calculation.

In both of these classes of models, simple enumeration of
meta-stable minimum is not a good account of the physical
probability distributions.

\section{Lessons from de Sitter's anti}

For completeness, I want to record here a number of results about
AdS space-times, which bear on the questions at hand.  In my
opinion, the most important result of the AdS/CFT correspondence is
that {\it the negative cosmological constant is a discrete input
parameter, which controls the high energy behavior of the theory. It
cannot be affected by low energy renormalization.}   In $AdS_5
\times S^5$ the cosmological constant in Planck units is $N^{1/4}$ .
It is determined by the rank of the gauge group.   This is not an
accident.  The coincidence of the AdS black hole entropy formula and
that of CFT, shows us that the c.c. in Planck units is always a
power of the c-coefficient in the CFT entropy formula.   Note that
this is completely consistent with low energy bulk effective field
theory.  In that formalism, the c.c. is an undetermined parameter,
which we have to fix by a renormalization condition.  AdS/CFT tells
us that, at least for negative c.c., this normalization can only be
understood in the high energy theory.  The effective field theory
must be tuned to reproduce that high energy input\footnote{In $1 +
2$ dimensional AdS, we can even describe this as forcing the low
energy effective theory to preserve the correct Virasoro symmetry
algebra of the high energy theory.}.

Everything we know about semi-classical AdS physics is consistent
with this.  There is no such thing as a ``meta-stable conformally
invariant state" in CFT.   Correspondingly, there are no tunneling
solutions from metastable points with any vacuum energy, into AdS
space.  The view of the c.c. in an AdS ``vacuum" as the minimum of a
low energy effective potential finds confirmation neither in
AdS/CFT, nor in semiclassical tunneling calculations.

AdS/CFT also gives us some clues about how the minimization of
effective potentials might be related to more robust concepts in
quantum gravity\cite{evas}.   The key idea is the {\it holographic
renormalization group}\cite{porratietal}, which reinterprets the
space-time equations for domain wall solutions to the
renormalization group flow in the boundary field theory.   Certain
minima and Breitenlohner-Freedman allowed maxima of the effective
potential of a supergravity theory, correspond to fixed points of
the RG flow.    Thus, the correct quantum gravitational notion of
defining a boundary conformal field theory can, in certain limits,
be related to finding stationary points of an effective potential.

There are a large number of puzzling features about this
correspondence, as well as features which indicate that conventional
ideas about the effective potential are unfounded.  I list some of
them, in random order:

\begin{itemize}

\item There are often stationary points of the potential which do
not appear to have any meaning in boundary field theory.   These
include points with tachyons that violate the BF bound.   These
could only be non-unitary CFTs, but they appear in the flow
initiated by relevant perturbations of a perfectly unitary theory.
Although there are no examples with meta-stable, positive energy,
minima, their possible role in the holographic RG is completely
obscure.

\item The holographic RG makes it clear that different stationary
points of the same potential do not represent different states of
the same theory.  The theory defined at the IR end of the flow has
fewer fundamental degrees of freedom than the UV theory.  It is
defined by decoupling.   In global dS space, the flow is cut off by
the compact sphere, and we never really lose the UV degrees of
freedom.   The perturbed theory has the same asymptotic value of the
c.c. as the unperturbed one.  The global space-time picture of these
domain walls is that of stable defects imbedded in a given AdS
space-time, rather than walls between two different asymptotic
space-times.

\item  There are domain walls in the low energy effective theory,
between positive curvature AdS stationary points, which have no RG
flow interpretation, because the CFTs have no relevant operators.

\item   The RG interpretation of the effective potential suggests
that only the behavior near some of its stationary points has
physical relevance.   It is well known that the RG has a vast {\it
scheme dependence} which makes most aspects of the RG flow simply
conventions.  It is only in the presence of large symmetry groups
(integrable flows) that there is physical relevance to the behavior
of the RG flow far from fixed points.  Note that the effective
potential, away from the stationary points determining the ends of
the flow, would then have a similar scheme dependence.

\end{itemize}

There is a very specific puzzle posed by the attempt to make the
landscape compatible with AdS/CFT.   In Type IIB string
compactifications with fluxes up to order $N$ on manifolds with
third Betti number $b_3$, one claims to find stable AdS minima with
c.c. as small as $N^{ - b_3}$.   This poses three problems:  one
must find CFTs corresponding to the AdS space with radius much
larger than both the Planck and the string scale.  Furthermore, it
is claimed that all of these vacua are in the region where ``string
perturbation theory" is a good approximation.

The AdS radius in Planck units is a measure of the $c$ coefficient
in the $2 + 1$ CFT entropy formula $S(E) = c (ER)^{2/3}$.   In $2 +
1$ dimensions we would expect any CFT to be in the IR basin of
attraction of a RG flow from a Gaussian fixed point, and we would
expect to see at least as much entropy at the Gaussian point as we
do at the fixed point representing the putative landscape state.
Silverstein\cite{eva} has suggested that the relevant Gaussian fixed
points are the world volume theories of branes carrying the fluxes
in the string compactification.   The entropy of these theories
appears to scale like $b_3 N^2$ rather than $N^{b_3}$. Silverstein
suggested that string junctions should (for reasons that are not
immediately apparent) be included among the UV degrees of freedom in
the world volume gauge theory.   These transform in the
$\otimes^{b_3} N_i$ under the $\prod U(N_i )$ gauge group of the
branes, and could provide the requisite entropy.

Gauge theories with large representations do have a large number of
fixed points in $2+1$ dimensions, which are easily accessible.  The
classical running of the gauge and quartic couplings can be balanced
by the large quantum correction coming from loops of the field in
the large representation.   The usual vector large $N$ resummation
gives a reliable calculation of the properties of these fixed
points.   The problem is that, since the anomalous dimensions at
these points are close to those of the Gaussian model, a large
radius interpretation of the CFT spectrum is not possible.

This is a general point which is worth emphasizing.   The number of
primary operators of fixed dimension grows exponentially with a
power of the dimension in any CFT which is an order one deformation
of a Gaussian model.   This is much too rapid to be explained by
Kaluza Klein states in a large radius $AdS_k \times K_p$
compactification.   In ${\cal N} = 4$ Super Yang-Mills theory with
$g^2 N$ of order $1$, we recognize this as an indication that ``the
AdS radius is of order the string scale", and attribute the growth
of the spectrum to excited string states.   Here this is not
possible.   We are not in a regime where the planar diagram
expansion makes sense.  Indeed, the effect of the huge
representation we invoked to explain the entropy is to make $g^2 N
\ll 1$ at the fixed point.

The landscape predicts a plethora of AdS flux vacua at weak string
coupling, with AdS radius ranging from $\sim 10$ to $N^{b_3}$ times
the string scale.   The CFT duals to these states have not yet been
found.  The search for them leads to a large number of $2 + 1$ CFTs
which have no dual interpretation in terms of large radius SUGRA, or
weakly coupled string theory on manifolds whose size is of order
string scale.   It seems to me that the continued search for the CFT
duals of AdS landscape states is the most likely avenue for finding
a rigorous justification for part of the landscape picture or for
falsifying it.

\section{Phenomenology of the landscape}

For practical purposes , the landscape gives us a large set of
alternative effective Lagrangians for describing the physics we have
observed or will observe in our universe.   These are parametrized
by a collection of numbers, which include the dimension of
space-time, the name, rank and representation content of the low
energy gauge theory, the value of the cosmological constant, and the
values of all the coupling constants and masses of fields in the
Lagrangian\footnote{In principle we could also have non-trivial
conformal field theories in the low energy world, at least in some
approximation.}.   These numbers can be collected together and
viewed as a multidimensional probability space.   In the
supergravity approximation, we have a way of calculating an {\it a
priori} distribution for these numbers.   Proponents of the
landscape would claim that this is an approximation to some more
exact distribution, though no-one has suggested a procedure for
calculating the corrections.    In the previous section I have
suggested that early universe cosmology may make important
modifications of the distribution of metastable minima. If it turned
out that the distribution predicted the Lagrangian we observe with
high probability, it would be a great triumph for string theory.

 The
value of the cosmological constant tells us that this is not the
case.   Weinberg's bound\cite{wein},which constrains cosmological
parameters by insisting on the existence of galaxies, has the form
(in Planck units).

$${\Lambda \leq K \rho_0 Q^3},$$
where $\rho_0$ is the dark matter density at the beginning of the
matter dominated era, $Q$ is the amplitude of primordial density
perturbations at horizon crossing, and K is a pure number of order
$1$.. For any reasonable values of the other parameters, this means
$\Lambda$ is much smaller than the typical value found in the
landscape.

It is clear then that we must supply additional data from experiment
in order to fix our description of the world.  The landscape
framework supplies some theoretical guidance - it tells us that
there are a finite number of possibilities, of order $10^{10^2}$ (to
order of magnitude accuracy in the logarithm). Various authors
\cite{douglasetal} have begun to investigate the {\it a priori}
distribution of properties like the gauge group and number of
generations, the scale of supersymmetry breaking, the existence of
large warp factors which give rise to large hierarchies of energy
scales {\it etc.}, assuming a uniform distribution on the space of
minima. The hope is that correlations will become evident which will
tell us that a small number of inputs is enough to extract the
Lagrangian of the world we live in from the ensemble of Lagrangians
the landscape presents us with.

The anthropic principle has also been invoked as an input datum to
impose on the ensemble of Lagrangians.  For its proponents, the
attraction of this principle is that the answer to a single yes/no
question, ``Is there carbon based life?" puts strong constraints
on a collection of parameters in the Lagrangian (assuming all
others fixed at their real world values).   This attraction may be
an illusion.  In a probability space, the characteristic function
of any subset of data points is a single yes/no question.   So
physicists must ask if there is any special merit to the
particular characteristic function chosen by the anthropic
principle.  This is a hard question to answer, because we do not
have much theoretical understanding of life and intelligence, and
we have no experimental evidence about other forms of life in the
universe we inhabit.

If the typical life form resembles the great red spot on Jupiter
rather than us, then this life form would think that the criterion
that is most appropriate to apply in our universe is the
Redspotthropic principle.   To put this in a more positive manner:
if considerations of carbon based life lead to explanations of the
values of the fundamental parameters, then we are making a
prediction about the typical form of life that our descendants will
find when they explore the universe.   It should look just like us,
or at least be sufficiently similar that the criteria for its
existence are close to the criteria for ours. If instead, our
descendants' explorations show that the typical life form could
tolerate much larger variations in the fundamental constants, then
our so-called explanations would really be a fine-tuning puzzle. The
Red Spot People could calculate and understand that {\it we}
wouldn't be there if the up quark mass were a little bigger, but
they might reasonably ask ``Who ordered them?".

Of all {\it soi disant} anthropic arguments, Weinberg's bound on
cosmological parameters is the least susceptible to this kind of
criticism.   If there are no galaxies, there are no planets, no Red
Spots, no Black Clouds, perhaps no conceivable form of life. Many
physicists cite the numerical success of this bound as evidence that
anthropic reasoning may have relevance to the real world.   It is
important to realize that this numerical success depends on keeping
all other parameters fixed. Arkani-Hamed reported on unpublished
work with Dimopoulos, which showed that if both $\Lambda$ and $M_P$
(really the ratio of these parameters to particle physics scales,
which are held fixed) are allowed to vary subject only to anthropic
constraints, then the preferred value of $\Lambda$ is larger than
experimental bounds by many orders of magnitude. Similarly the
authors of \cite{gw} following \cite{tegrees} argued that if both
$\Lambda$ and $Q$ are allowed to vary then the probability of
finding a universe like our own is of order $10^{-4}$. A contrary
result was reported in \cite{vil}, but only by assuming an {\it a
priori} probability distribution that favored small values of $Q$.

In inflationary models of primordial fluctuations, the value of $Q$
depends on details of the inflaton potential at the end of slow
roll.   We would certainly expect this parameter to vary as we jump
around the landscape.  Similar remarks apply to $\rho_0$. Allowing
$\rho_0$ to vary would further reduce the probability that the
anthropic distribution favors the real world. Thus, at least with
our current knowledge of the landscape, it seems likely that the
numerical success of Weinberg's bound in the landscape context is
not terribly impressive\footnote{I cannot resist remarking that in
the context of Cosmological Supersymmetry Breaking\cite{tbfolly},
only $\Lambda$ varies, and Weinberg's bound retains its original
numerical status.}.

The greatest challenge to all methods of dealing with the
landscape is the large number of parameters in the standard model
which have to be finely tuned to satisfy experimental constraints.
These include the strength of baryon, lepton and flavor violating
couplings, $\theta_{QCD}$ and the values of many quark and lepton
masses.    Anthropic reasoning helps with some of these
parameters, but not all, and is insufficient to explain the
lifetime of the proton, the value of $\theta_{QCD}$ and many
parameters involving the second and third generation quarks and
leptons. From the landscape point of view, the best way to deal
with this (in my opinion) is to find classes of vacua in which all
of these fine tuning problems are solved, perhaps by symmetries.
One can then ask whether there are enough vacua left to solve the
cosmological constant problem.   This might be a relatively easy
task.   One could then go on to see whether other features of this
class of vacua are in concordance with the real world.

It is clear that at a certain point in this process, if we don't
falsify the landscape easily\footnote{{\it e.g.} by showing that the
number of vacua left after all the other fine tuning problems are
solved is too small to solve the cosmological constant problem.}, we
will run into the problem that current technology does not allow one
to calculate the low energy parameters with any degree of precision.
Indeed, the error estimates are only guesses because we don't even
know in principle how to calculate the next term in the expansion in
large fluxes.    A more fundamental framework for the discussion of
the landscape is a practical necessity as well as a question of
principle.  I have suggested that if a rigorous framework for the
landscape exists, it is probably to be found in the context of a
theory of a Big Bang universe with Eternal Inflation, and future FRW
asymptotics for any given observer. It is likely that we will be
unable to define more precise calculations of the properties of the
landscape without finding a rigorous mathematical definition of such
a space-time.

The fundamental object in such a space-time would be a scattering
matrix\cite{mcosmo} relating a complete set of states at the Big
Bang to states in Lorentzian CDL bubble space-times corresponding to
decays of meta-stable dS landscape states into the Dine-Seiberg
region of moduli space.   The final states, in addition to particle
labels, would carry indices ,$(V,L)$ describing a particular dS
minimum $L$ and a particular ``asymptotic vacuum", $V$, into which
it decays.   Thus, we would have matrix elements
$$S(I| V,L, p_i),$$ where $I$ labels an initial state at the Big
Bang and $p_i$ a set of ``particle" labels for localized scattering
states in a given CDL bubble.   An important unanswered question is
whether the S-matrix is unitary for each $(L,V)$\footnote{One would
then invoke the existence of unitary mappings taking the different
unitary S-matrices into each other.} or only when all $(L,V)$
sectors are taken into account. The first alternative is analogous
to the proposal of \cite{susspriv}.

The following argument has a bearing on this question.   I have
stated above that there was no problem with an infinite number of
final states for fixed $L$.   This is not necessarily the case.   If
I extrapolate scattering data on ${\cal I}_+$ backwards, using the
classical equations of motion, and assuming a minimal finite energy
for each particle, then all but a finite number of states will
encounter a space-like singularity before transition to the
metastable dS regime.   This is the time reverse of the argument
about black holes in the initial state of the time symmetric CDL
bubble.  This suggests that for fixed $L$, the matrix $S(I| V.L, p_i
)$ has finite rank: only a finite subspace of the space of all out
states on the CDL geometry labeled by $L,V$ would be allowed. This
leads to a modification of the proposal of \cite{susspriv} in which
only the S-matrix for fixed $V$, keeping all possible values of $L$,
is unitary. However, there is no clear reason now to assume that $V$
should be fixed, so perhaps only the full S-matrix is unitary.

A more disturbing conclusion is reached if one combines the claim of
\cite{mrd} that the number of $L$ sectors is finite, with the above
argument.   One then concludes that the whole subsector of the
S-matrix that has meta-stable dS resonances, also has finite rank,
and that the entire landscape fits into a Hilbert space with a
finite number of states. There will be an infinite rank matrix of
direct transitions between the Big Bang and a final FRW state, which
do not go through meta-stable dS resonances. One of the supposed
virtues of the landscape picture of metastable dS was that, unlike a
stable dS space, the landscape was part of a system with an infinite
number of states, which could make infinitely precise quantum
measurements on itself. If both Douglas' claim, and that of the last
paragraph are true, this is no longer obvious. All but a finite
number of the final states in a given CDL instanton geometry, would
not connect to a tunneling process from a meta-stable dS vacuum, but
instead would evolve directly from a Big Bang.    The whole issue of
a rigorous framework for the landscape remains as murky as ever.

It seems to me that the obvious conclusion to draw from the
conjunction of the difficulties of principle from which the
landscape suffers, with its phenomenological problems, is that the
landscape does not exist.  This is not meant to be a firm scientific
conclusion, but a warning and a guide to future research.   In my
opinion, much of the justification for the landscape relies on
``ancient string theory folklore", which has been discredited by the
insights of the duality revolution, most particularly the AdS/CFT
correspondence.  Careful consideration of classical black hole
physics also casts doubt on the world view of the landscape
gardener.

At times the landscape is also hailed as the only way to understand
de Sitter space-time in string theory.   A (formerly? )famous
footnote in the field theory textbook by Bjorken and Drell, warns
the reader to be careful of making the ``what else could it be?"
argument, because, ``if you're not careful, someone else will tell
you what else it could be".  In this case, perhaps, someone already
has \cite{tbfolly}.

\section{\bf Acknowledgements}

I would like to thank L. Baulieu for inviting me to the Cargese
workshop, and Nima Arkani-Hamed for conversations which contributed
to my understanding of the issues discussed here.

This research was supported in part by DOE grant number
DE-FG03-92ER40689.


\begin{thebibliography}{19}

\bibitem{isovac} T.~Banks, {\it On Isolated Vacua and Background Independence},
hep-th/0011255.
\bibitem{hetero} T.~Banks, {\it A Critique of Pure String Theory: Heterodox Opinions of
Diverse Dimensions}, hep-th/0306074.
\bibitem{ss} N.~Seiberg, S.~Shenker, {\it A Note on Background
(In)dependence}, Phys. Rev. D45, 4581, (1992), hep-th/9201017.
\bibitem{gf} E.~Farhi, A.~Guth, {\it An Obstacle to Creating a
Universe in the Laboratory}, Phys. Lett. B183, 149, (1987).
\bibitem{cdl} S.~Coleman, F.~DeLuccia, {\it Gravitational Effects On
and Of Vacuum Decay}, Phys. Rev. D21, 3305, (1980).
\bibitem{susspriv}B.~Freivogel, L.~Susskind, {\it A Framework for
the Landscape}, hep-th/0408133.
\bibitem{tHsuss} C.R.~Stephens, G.~'t Hooft, B.F.~Whiting, {\it Black Hole Evaporation
Without Information Loss}, Class. Quant. Grav. 11:621, (1994);
L.~Susskind, L.~Thorlacius, J.~Uglum, {\it The Stretched Horizon and
Black Hole Complementarity}, Phys. Rev. D48, 3743, (1993).
\bibitem{mcosmo} T.~Banks, W.~Fischler, {\it M-theory Observables
for Cosmological Space-times}, hep-th/0102077.
\bibitem{moorehorne} G.~Moore, J.~Horne, {\it Chaotic Coupling Constants},
Nucl. Phys. B432,
109, (1994);T.~Banks, M.~Berkooz, G.~Moore, S.~Shenker,
P.J.~Steinhardt, {\it Modular Cosmology}, Phys. Rev. D52, 3548,
(1995).
\bibitem{tbf} T.~Banks, W.~Fischler, {\it An Holographic Cosmology}, hep-th/0111142;
{\it Holographic Cosmology 3.0}, hep-th/0310288; T.~Banks,
W.~Fischler, L.~Mannelli, {\it Microscopic Quantum Mechanics of the
$p=\rho $Universe}, hep-th/0408076.
\bibitem{wein} S.~Weinberg, {\it Anthropic Bound on the Cosmological
Constant}, Phys. Rev. Lett. 59, 2607, (1987).
\bibitem{douglasetal} M.R.~Douglas, {\it Statistics of String
Vacua}, hep-ph/0401004; F.~Denef, M.R.~Douglas, {\it Distributions
of Flux Vacua}, JHEP 0405:072, 2004, hep-th/0404116; F.~Denef,
M.R.~Douglas, B.~Florea, {\it Building a Better Racetrack}, JHEP
0406:034, 2004, hep-th/0404257; M.R.~Douglas {\it Statistical
Analysis of the Supersymmetry Breaking Scale}, hep-th/0405279; {\it
Basic Results in Vacuum Statistics}, hep-th/0409207; A.~Giryvaets,
S.~Kachru, P.K.~Tripathy, {\it On the Taxonomy of Flux Vacua}, JHEP
0408:002,2004, hep-th/0404243; O.~DeWolfe, A.~Giryvaets, S.~Kachru,
W.~Taylor, {\it Enumerating Flux Vacua With Enhanced Symmetries},
hep-th/0411061
\bibitem{gw} M.L.~Graesser, S.D.H.~Hsu, A.~Jenkins,M.~Wise,
{\it Anthropic Distribution for Cosmological Constant and Primordial
Density Perturbations}, Phys. Lett. B600, 15, (2004).
\bibitem{tegrees} M.~Rees, M.~Tegmark, {\it Why is the CMB Fluctuation Level $10^{-5}$},
Ap. J. 499, 526, (1998).
\bibitem{vil} L.~Pogosian, M.~Tegmark, A.~Vilenkin, {\it Anthropic
Predictions for Vacuum Energy and Neutrino Masses}, JCAP 0407:005,
2004, astro-ph/0404497.
\bibitem{tbfolly} T.~Banks, {\it Cosmological Breaking of
Supersymmetry?}, hep-th/0007146; and Int. J. Mod. Phys. A16, 910,
(2001).
\bibitem{mrd} M.R.~Douglas, {\it The Statistics of String/M-theory
Vacua}, JHEP 0305:046,2003, hep-th/0303194.
\bibitem{eva}E.~Silverstein, {\it AdS and dS Entropy From String
Junctions}, hep-th/0308175.
\bibitem{porratietal} L.~Girardello, M.~Petrini, M.~Porrati, A.~Zaffaroni,
JHEP 9812, 022,(1998), hep-th/9810126; D.Z.~Freedman, S.S.~Gubser,
K.~Pilch, N.P.~Warner, Adv. Theor.Math.Phys.3, 363, (1999),
hep-th/9904017.
\end{thebibliography}
\end{document}